# Quantitatively Designing Porous Copper Current Collectors for Lithium Metal Anode

Bingyu Lu[1†], Edgar Olivera[2, 3†], Jonathan Scharf[1], Mehdi Chouchane[5,6], Chengcheng Fang[1,10], Miguel Ceja[1], Lisa Pangilinan[3], Shiqi Zheng[2,4], Andrew Dawson[3], Diyi Cheng[7], Wurigumula Bao[1], Oier Arcelus[5,6], Alejandro A. Franco[5, 6,8,9], Xiaochun Li[2,4], Sarah H. Tolbert[2, 3*], Ying Shirley Meng[1*]

[1]Department of NanoEngineering, University of California San Diego, La Jolla, CA 92093, USA

[2]Materials Science and Engineering, University of California Los Angeles, Los Angeles, CA, 90095, USA

[3]Chemistry and Biochemistry, University of California Los Angeles, Los Angeles, CA, 90095, USA

[4]Mechanical and Aerospace Engineering, University of California Los Angeles, Los Angeles, CA, 90095, USA

[5]Laboratoire de Réactivité et Chimie des Solides (LRCS), UMR CNRS 7314, Université de Picardie Jules Verne, Hub de l'Energie, 15 Rue Baudelocque, 80039, Amiens, France

[6]Réseau sur le Stockage Electrochimique de L'Energie (RS2E), FR CNRS 3459, Hub de l'Energie, 15 Rue Baudelocque, 80039, Amiens, France

[7]Materials Science and Engineering Program, University of California, San Diego, La Jolla, CA 92121, USA

[8]ALISTORE-European Research Institute, FR CNRS 3104, Hub de l'Energie, 15 Rue Baudelocque, 80039, Amiens, France

[9]Institut Universitaire de France, 103 Boulevard Saint Michel, 75005, Paris, France

[10]Department of Chemical Engineering and Materials Science, Michigan State University, 220 Trowbridge Rd, East Lansing, MI 48824, USA.

*Correspondence to: shirleymeng@ucsd.edu, tolbert@chem.ucla.edu

†These authors contributed equally

**Abstract:** Lithium metal has been an attractive candidate as a next generation anode material. Despite its popularity, stability issues of lithium in the liquid electrolyte and the formation of lithium whiskers have kept it from practical use. Three-dimensional (3D) current collectors have been proposed as an effective method to mitigate whiskers growth. Although extensive research efforts have been done, the effects of three key parameters of the 3D current collectors, namely the surface area, the tortuosity factor, and the surface chemistry, on the performance of lithium metal batteries remain elusive. Herein, we quantitatively studied the role of these three parameters by synthesizing four types of porous copper networks with different sizes of well-structured micro-channels. X-ray microscale computed tomography (micro-CT) allowed us to assess the surface area, the pore size and the tortuosity factor of the porous copper materials. A metallic Zn coating was also applied to study the influence of surface chemistry on the performance of the 3D current collectors. The effects of these parameters on the performance were studied in detail through Scanning Electron Microscopy (SEM) and Titration Gas Chromatography (TGC). Stochastic simulations further allowed us to interpret the role of the tortuosity factor in lithiation. By understanding these effects, the optimal range of the key parameters is found for the porous copper anodes and their performance is predicted. Using these parameters to inform the design of porous copper anodes for Li deposition, Coulombic efficiencies (CE) of up to 99.56% are achieved, thus paving the way for the design of effective 3D current collector systems.

**Main text:** With a high theoretical capacity (3,860 mAh/g, or 2,061 mAh/cm$^3$) and low electrochemical potential (–3.04 V versus the standard hydrogen electrode), lithium metal is considered as the ideal candidate for the next generation battery anodes.[1] In fact, the lithium metal anode is a key component of next-generation high-energy-density battery systems such as lithium-sulfur and lithium-oxygen batteries.[2] However, there are many formidable obstacles that need to be solved before lithium metal anodes can be effectively used in commercial cells. The fundamental problem lies in the formation of lithium whiskers during plating, which eventually leads to the formation of inactive lithium after cycling.[3] As a result, the lithium metal anode suffers from low CE and low cyclability.

Numerous methods have been proposed to mitigate the lithium whisker formation issue and improve the cyclability of lithium metal anode. One of the most promising methods to achieve this is through the engineering of the electrolyte,[4] such as adding additives,[5–7] using high lithium salt concentration[8–10] and localizing high lithium-salt concentration.[11–13] The main working principle of a high-performance electrolyte is to construct a homogeneous SEI layer and provide a uniform lithium ion flux during the plating process so that a smooth and dense morphology can be achieved.[14] Another emerging methods is the engineering of current collectors. The current collectors play a crucial role in the performance of a lithium metal battery cell. Planar Cu foils have been used as the anode current collector for decades because of their electrochemical stability against lithium.[15] However, under practical current densities, due to the inhomogeneous lithium-ion flux, lithium whiskers can easily form on the planar Cu foils during cycling.[16] Therefore, a current collector that can regulate the local current density and provide a uniform lithium ion flux is desired for lithium metal anodes.[15, 17] A variety of 3D current collectors have been designed to achieve this purpose.[18–24]

For a high-performance 3D current collector, there are three key parameters that need to be carefully designed: surface area, tortuosity factor, and surface chemistry.[25] However, these three parameters are correlated with each other, and often one cannot be altered without disrupting the other two. Several works have mentioned the effects of these key parameters on the performance of the 3D current collectors.[26, 27] Yun *et al.* discovered that by tuning the dealloying time of brass foil, the resulting copper foil would have different pore sizes and surface areas, which eventually led to different performance of the dealloyed copper foils.[26] Similarly, by tuning the Cu pillar size and spacing, Chen *et al.* were able to study effects of surface area and pillar spacing on the CE of the Cu pillar current collector and found the best combination for the performance.[27] Therefore, there is no doubt that surface area, tortuosity factor, and surface chemistry play crucial roles in determining the performance of the 3D current collectors. However, quantitative analysis of these key parameters is required when designing a new 3D current collector system.

In the present work, 3D porous Cu current collectors were fabricated by etching Fe from Cu-Fe composites with different compositions. The physical properties of current collectors were then quantified using laboratory micro-CT,[28,29] and their performance was predicted in terms of surface areas, pore sizes and tortuosity factors. The quantitative study suggested that the high surface area was not as beneficial as previously believed[30, 31] and that the tortuosity factor should always be kept at a minimum. The high performance can only be achieved in a system when the three key parameters are in its optimal range. The prediction of the performance was also validated by Titration Gas Chromatography (TGC)[32] and electrochemical testing.

**Experimental**

1) Synthesis of the porous copper networks

Pure Cu powders (Fisher Scientific, electrolytic powder) and pure Fe powder (Beantown Chemical, -325 mesh, reduced, 98%) were used to prepare Cu-Fe precursor ingots with three different compositions: 10 atomic percent (at%) Cu-90at%Fe, 20at%Cu-80at%Fe, and 30at%Cu-70at%Fe. The average powder size of both powders was about 30 microns. A hydraulic press was used to make pellets of 1 cm in diameter for each composition. These pellets were then arc melted five times to make homogeneous ingots with high cooling rates. In addition, the 30at%Cu-70at%Fe was loaded in a 3cm-diameter alumina crucible and heated in an Ar-filled furnace at 1550 °C for 1 hour, followed by slow cooling down to room temperature. The four different types of samples were sliced and polished using sandpaper to make samples 1 cm in diameter and 200 microns in thickness. These four samples are denoted throughout this paper as 10Arc, 20Arc, 30Arc and 30Furnace in accordance with their composition and synthesis method.

After the Cu-Fe ingots were successfully synthesized, the disk was then etched in 5wt% $H_2SO_4$ acid solution at 90°C under constant stirring for 24 hours. The acid solution was changed every 8 hours to prevent the accumulation of contaminants. The etched sample was then cleaned in 5wt% HCl solution under sonication for 15 minutes. This was followed by an acetone wash under sonication for 15minute to wash away any leftover contaminants. Finally, the cleaned samples were dried under vacuum for one hour.

2) Zn Electrodeposition

The 20Arc Cu sample was Zn-coated using an electrolyte of 0.05 M $ZnSO_4$ +0.3M $H_3BO_3$ and the 30Furnace Cu sample was Zn-coated using an electrolyte of 0.01 M $ZnSO_4$ +0.3M $H_3BO_3$. In both cases, a small amount of 1M $H_2SO_4$ was added to adjust the pH to 1-2. The cleaned porous Cu was removed from the glovebox and used as the working electrode in a three-electrode set up where the counter electrode was platinum foil and the reference electrode was

Ag/AgCl in KCl. A current density of 1mA/cm$^2$ was used to deposit Zn for 10 min and 5 min for the 30Furnace and 20Arc, respectively. The area used for the deposition current density was the measured effective surface area. The sample was then washed with deionized water and acetone and then dried under vacuum for one hour.

3) Electrochemistry

CR2016-type coin cells were assembled in an Ar-filled glove box for electrochemical characterization. The electrolyte consisted of 75 µL of 1 M lithium bis(trifluoromethane sulfonyl)imide (LiTFSI) in a mixed solvent of 1,3-dioxolane (DOL) and 1,2-dimethoxyethane (DME) (1:1 in volume) with 2% LiNO3 and Li metal foil was used as the counter electrode. A range of current density was used during lithium plating morphology study.

4) Scanning Electron Microscope / Energy Dispersive X-ray Spectroscopy

A FEI Apreo Scanning Electron Microscope (SEM) was used to study the structure of the as-prepared porous copper current collector as well as the morphology of the electrochemically deposited lithium (EDLi). Energy Dispersive X-ray Spectroscopy (EDS) was used to characterize the elemental composition of the sample.

5) Micro-CT

The samples were individually punched into films with a 2 mm radius piece and were stacked in a PTFE cylindrical tube with alternating PTFE films to provide separation. Two scans were conducted using a ZEISS Xradia 510 Versa micro-CT instrument. The first scan was conducted on the lager pore samples (30Furnace and 10Arc, shown in Fig 1c-d) and had a voxel size of 1.07 µm and an exposure of 4 seconds. The second scan was used to examine the smaller pore samples (20Arc and 30Arc, shown in Fig 1a-b) and had a voxel size of .7834 µm and an exposure of 6 seconds. Both scans were conducted with 1801 projections at an X-ray energy of 140 keV with a 71.3 uA current using a high energy filter at a 4X magnification. The beam hardening constant for the reconstruction was 0 and 0.6 for the first and second scan respectively. Post measurement analysis was performed by the Amira-Avizo method using the Deblur, Delineate, and Median Filter modules for data sharpening and filtration provided by the software. The Separate Objects module, which utilizes a distance map and a watershed algorithm, was used to define 3D pores and determine the overall distribution of the pore sizes. The tortuosity factor was determined for each of the resulting structures using the software TauFactor[33] and cross-validated with GeoDict, both relying on Fickian diffusion.

Simulation of the Li deposition process in the porous copper made use of an *in house* MATLAB-based stochastic algorithm. For a given capacity, the volume of deposited Li was determined considering 100 % coulombic efficiency and a molar volume of $13.02 \times 10^{-6}$ m$^3$/mol. Then, this volume was converted into a number of pixels. Subsequently, pixels of void in contact with solid phase (Cu or Li) were converted into pixels of Li until the desired amount was achieved. To mimic the experimental observations, a percentage of the Li was deposited on the top of the copper support (see Table S1). We arbitrarily allowed for an extra 15 µm thick void to be inserted on top of the current collector for Li deposition. After Li deposition, the remaining empty thickness was removed from this extra 15 µm space in order to limit its impact on the tortuosity factor calculation. Additionally, to account for the heterogeneous deposition of Li along the thickness, an arbitrary gradient has been applied to replicate the experimental observations. Each structure was divided into 6 equal sub-volumes for which the deposited Li amount is reported in Table S2. Then the algorithm called the software TauFactor[33] to determine the tortuosity factors of each porous copper network. Each condition was repeated between 5 and 10 times to quantify the uncertainty linked to the stochastic method.

6) Titration Gas Chromatography (TGC)

The TGC method was used to quantify the amount of inactive metallic lithium formed in the porous copper after cycling. After plating and stripping of Li, the porous Cu electrode was recovered from the coin cell and the porous copper film, including any residual inactive lithium, together with the separator, were put into a 30 mL bottle without washing. The bottle is then sealed with a rubber stopper and the internal pressure of the bottle was adjusted to 1 atm. After removing the bottle from the glovebox, excess deionized (DI) water (0.5 mL) was injected into the bottle to react with any residual inactive metallic lithium to form $H_2$ gas. The vial was then well mixed by shaking, and a gas-tight syringe is used to quickly take 30 µL of the gas from the head-space of the sealed bottle. The gas was then injected into a Nexis GC-2030 Gas Chromatograph (Shimadzu) for $H_2$ quantification. A pre-established $H_2$ calibration curve was used to calculate the amount of inactive metallic lithium from the measured $H_2$ peak area. The mass of inactive metallic lithium in the porous copper films was directly related to the amount of $H_2$.

7) X-Ray Diffraction (XRD)

XRD was performed using a PANalytical X'Pert Pro powder diffractometer operating with Cu Kα radiation ($\lambda$ = 1.5418 Å) using a 0.03° step size, a voltage of 45 kV, and a current of 40 mA. XRD patterns were recorded in the range of 15° < 2θ < 85°.

8) X-Ray Photo-electron Spectroscopy (XPS)

XPS analysis was performed using a Kratos Axis Ultra DLD spectrometer with a monochromatic Al (Kα) radiation source. A charge neutralizer filament was used to control charging of the sample, a 20 eV pass energy was used with a 0.1 eV step size; scans were calibrated using adventitious carbon by setting the C 1s peak to 284.8 eV. Samples were etched with an Ar beam with a raster size of 2mm x 2mm at an energy of 4 kV for 1 minute.

**Result and Discussion**

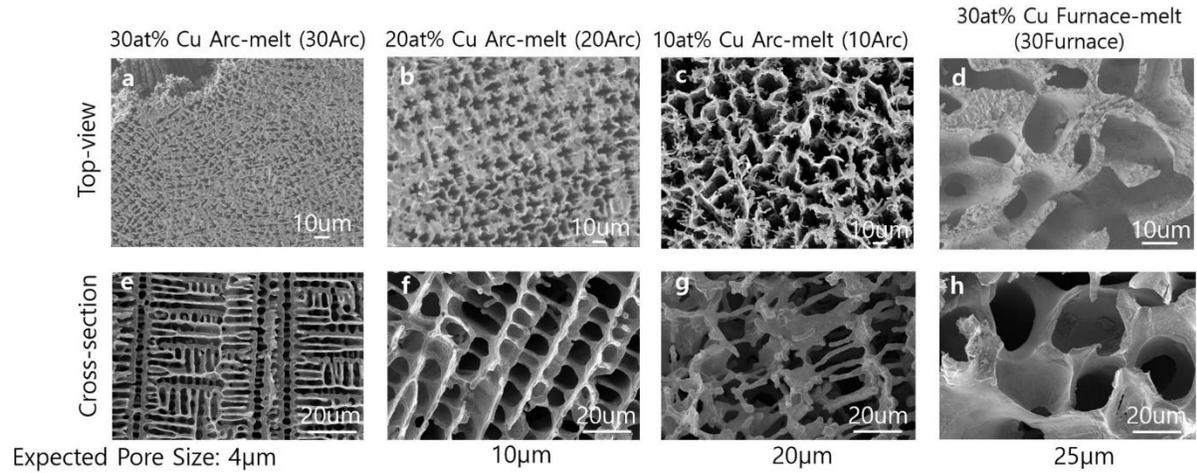

Figure 1. SEM images of the four porous copper films used in this work. a) Top-view and e) cross-section of the 30 at% Cu arc-melted (30Arc) sample; b) top-view and f) cross-section of the 20 at% Cu arc-melted (20Arc) sample; c) top-view and g) cross-section of the 10 at% Cu arc-melted (10Arc) sample; d) top-view and h) cross-section of the 30 at% Cu furnace-melted (30Furnace) sample.

Porous copper films with interconnected channels were fabricated by etching Fe from Cu-Fe composites. Ingots with compositions of 10, 20 and 30 atomic percentage (at%) Cu were made by arc melting and one ingot with a composition of 30 at% Cu was made using a conventional Ar-filled furnace. Arc melting involves rapid heating and rapid cooling, while heating and cooling rates in a conventional furnace are much longer. Because Cu and Fe are essentially immiscible in the solid state but fully miscible in the melt, the size of the phase separated domains that form upon solidification are strongly dependent on the cooling rate, with faster cooling of the arc furnace producing smaller domains. The XRD patterns of the as-made ingots are shown in the SI in Figure S1 and are in agreement with a phase separated mixture of Cu and Fe. For a given cooling rate, the size of domains can be further tuned using the atomic percentage of Cu and Fe with higher Fe fractions resulting in large Fe domains. Because the Fe is the fraction that is etched form the ingot, both larger Fe fractions and slower cooling rates result in larger pore in the remaining Cu. After careful etching and cleaning, porous coppers with different well-structured channels were fabricated. (Fig. 1). As observed in the XRD patterns of the post-etching samples (Fig. S2), the porous copper samples obtained from arc-melted ingots are (200) oriented and the porous copper materials made using a conventional furnace are polycrystalline. Furthermore, the XPS data for all four samples (Fig. S3-S6) show that the samples are mainly composed of Cu and some remaining copper oxide on the surface. However, there is no Fe or S remaining in the samples. Unlike other randomly structured porous copper that were also synthesized by dealloying method,[26] the porous copper networks derived from Cu-Fe composites show homogeneous, large pored micro-structures that allow us to quantitatively analyze the effects of key parameters such as surface area, pore size, and tortuosity factor on the performance of a 3D current collector.

As shown in Fig. 1a and 1e, when 30 at% Cu was mixed with Fe and arc with rapid heating and cooling, the resulting porous copper had roughly 4µm-wide and 10µm-long channels, which provide empty spaces for lithium to deposit. As the Cu at% decreased in the precursor, the channels in the final porous Cu grew longer and wider. The 10 at% Arc melt porous copper showed approximately 20µm-wide channels (Fig. 1c, 1g), but the structure of the channels was very fragile. Therefore, a second melting technique was used when synthesizing the 25µm-wide porous copper, which involved slow heating and cooling an ingot with 30 at% Cu using an Ar-filled furnace. The resulting porous copper showed 25µm-wide pores with sturdy ligaments.

The physical parameters of the porous copper samples were quantified by laboratory X-ray micro-CT. Each piece of the porous copper was punched and fitted into a sample holder tube. With voxel size of 0.7834 µm or 1.07 µm, the 3D structures (Fig. 2a, 2b) of the porous coppers obtained through the tomography was analyzed using the Amira-Avizo software package and the physical properties of the porous coppers were measured (Fig. 2).

Fig. 2d reports the pore size distribution of each porous copper sample. As expected, the 30Furnace sample shows the largest pores and the 30Arc sample had the smallest pores, followed by 20Arc and 10Arc. The peak of each histogram curve represents the average length of the micro-channels in the porous copper. The results match well with the SEM images shown in Fig. 1. The geometric tortuosity values in Fig. 2e were calculated based on the average length a fictive particle needs to travel to go through the whole porous copper sample in the z-direction, which approximately quantifies how easy it is for a lithium ion to diffuse and electro-migrate inside the Cu network during an electrochemical plating process. Both the 10Arc and 30Furnace samples have relatively low geometric tortuosity because they have large enough pores for lithium ion to move freely without many obstacles in the z-direction. The volumetric surface area of each porous network can be directly obtained from the 3D tomography (Fig. 2c). The effective surface area was calculated based on the actual size of the electrode used in the electrochemical testing (7 mm in diameter and 200 µm in thickness). In the surface area measurement, the same trend is observed as in the geometric tortuosity: the samples with smallest pore sizes have the largest effective surface area while the larger pores give the smallest surface areas. There is a slow drop of effective surface area from the 20Arc to 10Arc materials, despite the fact that the pore size difference in these two samples is large. This is caused by the change of ligament size as the at% of Cu in the precursor ingot decreases from 20 to 10 (Fig. 1f, 1g). The thinner ligament contributes to the extra surface area in the 10Arc sample. With these key physical parameters quantified, the effects of morphology and the distribution of deposited Li can now be studied in detail.

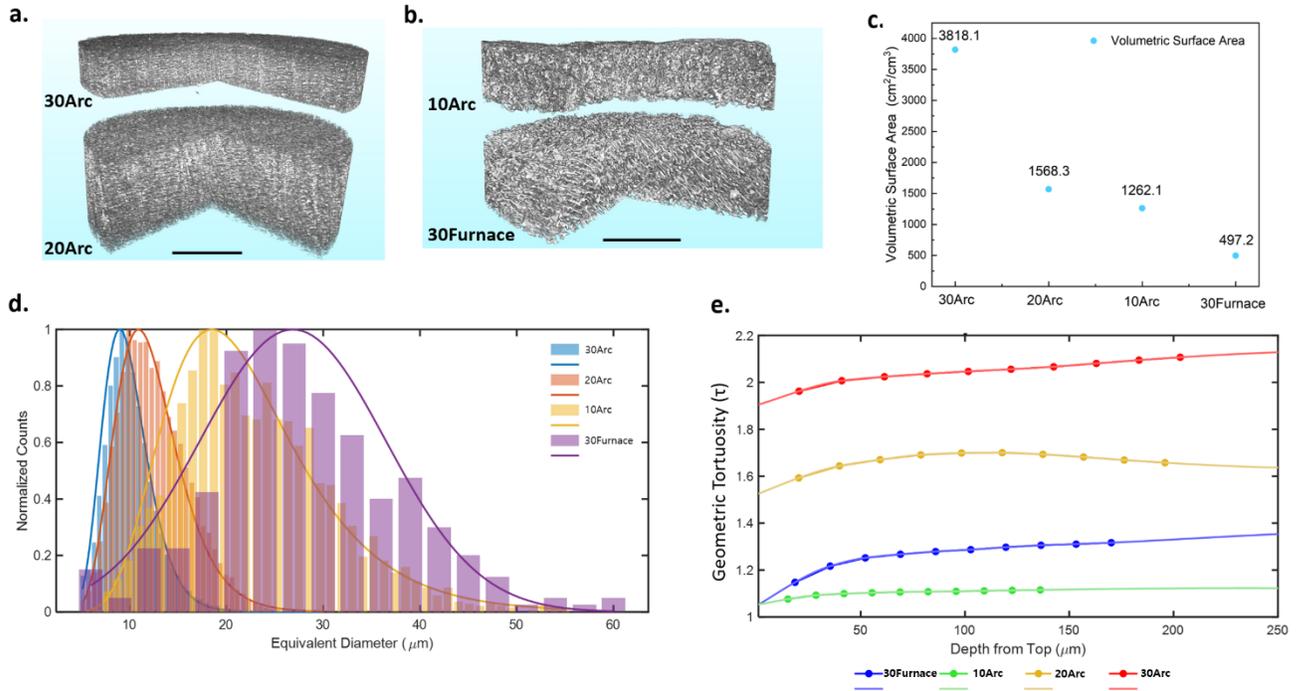

Figure 2. a) The 3D tomography of 30Arc and 20Arc with voxel size of 0.7834 µm. b) The 3D tomography of 10Arc and 30Furnace with voxel of 1.07 µm. For both parts a) and b), the scale bar is 200 µm. c) Volumetric surface area in the porous copper; d) Histograms of the pore size distribution for each porous copper sample. e) Geometric tortuosity of the porous copper as a function of depth.

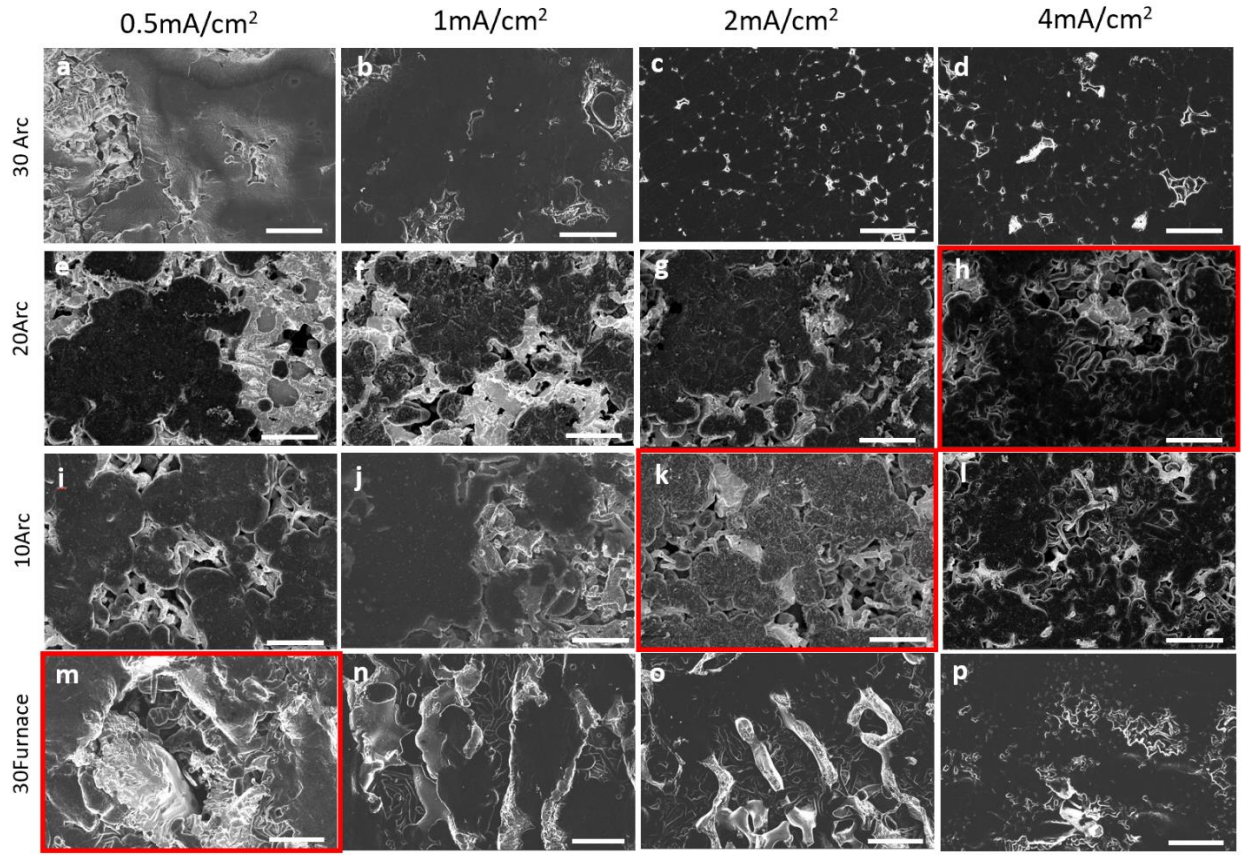

Figure 3. Top-view SEM images of the morphology of lithium plated under different current densities onto the porous copper networks with different pore sizes: a-d) 30Arc; e-h) 20Arc; i-l) 10Arc; m-p) 30Furnace. Current densities are indicated on the figure. The critical current density at which lithium whiskers start to grow are marked with red boxes. All samples were plated to 4 mAh/cm$^2$ before imaging.

The effect of the surface area on the morphology of the deposited lithium was characterized using top-view SEM images. For each sample, 4 mAh/cm$^2$ of lithium was plated onto the porous copper using 4 different current densities, each of which are indicated in Fig. 3. As the surface area decreases with the increase of pore size, the effective local current density per unit surface area of Cu would also increase. It is widely accepted that high local current densities can lead to an inhomogeneous flux of lithium ions during plating, causing the formation of lithium whiskers.[17] By studying the lithium deposition morphology on the porous Cu networks with different surface areas under a range of current densities, the critical current density at which lithium whiskers begins to grow in each sample can be determined.

With the largest effective surface area, which is 29.49 cm$^2$ in a piece with 0.7 cm diameter and 200 μm thickness, the 30Arc samples produces a dense and smooth lithium deposition morphology throughout the four current densities tested. As the effective surface area decreased to 12.03 cm$^2$, the 20Arc samples also show relatively dense morphology in most of the current densities tested except at 4 mA/cm$^2$, which is indicated by the red box in Fig. 3h. At this current density, lithium whiskers begin to grow on the surface of the pore channels, which means that the local current density becomes too high, inducing inhomogeneous lithium ions transport. As the effective surface area is further decreased, the critical current density also decreased. In the 30Furnace sample, where the effective surface area was only 3.76 cm$^2$, lithium whiskers were observed at current densities as low as 0.5 mA/cm$^2$, which means the critical current density might be even lower than that. Based on these observations alone, it seems that the higher the surface area in the porous copper, the better the Li morphology.

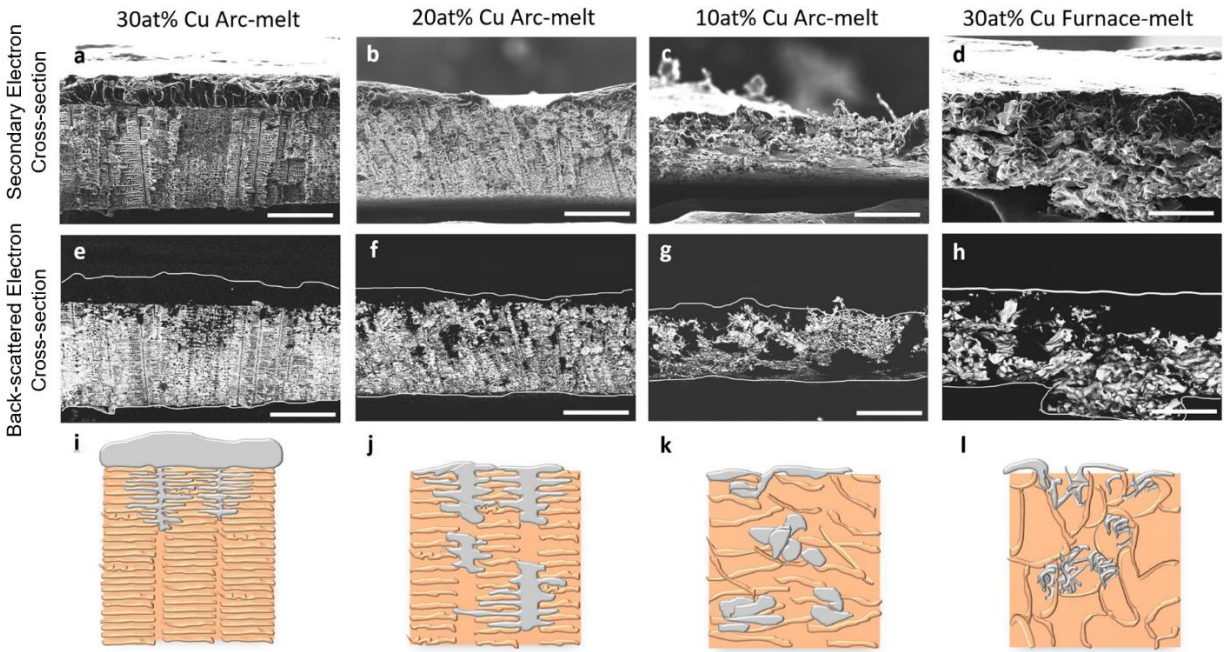

Figure 4. a-d) Cross-sectinoal secondary electron SEM images, e-h) back-scattered electron SEM images and i-l) cartoon illustrations of the observed lithium spatial distribution in the cross-section of each porous Cu sample.  All samples were plated to 20 mAh/ cm$^2$. The scale bars in all images correspond to 50 µm.

The effect of the geometric tortuosity and pore size of the porous Cu network on the deposited lithium morphology was studied by cross-sectional SEM images. Fig. 4 shows the distribution of deposited lithium in each porous Cu sample after plating for 20 hours at 1 mA/cm$^2$. Back-scattered electron (BSE) images (Fig. 4e-h) give a clear view of how lithium is distributed across the whole porous copper network: the brighter region is copper, the darker region is lithium, and the boundary between vacuum and lithium is marked by white lines. Cartoon schemes were also constructed to better illustrate the observed lithium deposition distribution and morphology in the different porous copper samples (Fig. 4i-l).

With the smallest pore diameter (~4 µm) and the highest z-directional geometric tortuosity (1.35), the 30Arc samples gave the most inhomogeneous distribution of the deposited lithium within the volume (Fig 4a, 4e, 4i). Most of the deposited lithium accumulated on the top surface on the porous copper, and the space inside the pores, where the lithium was supposed to deposit, was mostly empty. This inhomogeneous distribution of deposited lithium can be attributed to the complex structure of the porous copper, with small pores and high geometric tortuosity hampering lithium ions transport and accessibility to the majority of the structure pores. As we increased the pore size to ~10 µm, which led to a decrease in the z-directional geometric tortuosity to 1.28, the distribution of the deposited lithium in the porous copper changed dramatically (Fig 4b, 4f, 4j). More lithium was deposited inside the pores of the Cu instead of on the top surface. The higher utilization of the empty pores increases the contact area between lithium and the copper, which should lower the local effective current density and result in a more uniform morphology. The further increase in the pore size (~20 µm) and the decrease in the geometric tortuosity (1.03) allowed more lithium to grow inside the pores (Fig 4c, 4g, 4k). The deposited lithium grew in a bulky way and had a clustered morphology inside the porous Cu. In the largest porous Cu (~25 µm), the distribution of the deposited lithium was similar to that in the 10Arc sample. However, the increase of the pore size also led to a decrease of surface area. As a result, the morphology of the deposited lithium changed from bulky to whisker-like (Figure 4d, 4h, 4l). This change of morphology should eventually lead to the formation of inactive metallic lithium and cause the decrease in coulombic efficiency.[3]

To gain further insights on the factors limiting Li deposition in these porous copper structures, a computational study focused on the tortuosity factor was carried out. Such a study was designed to capture the evolution of the tortuosity factor with the increase of Li deposition in the different porous Cu structures. The tomographic structure of each sample was imported into MATLAB and the full whole volume of each sample was used for the calculation. An *in-house* MATLAB®-based algorithm was then used for the stochastic generation of lithium deposited in each porous copper sample. This algorithm favored the deposition of lithium in aggregated form, since that is what was observed experimentally. To further match with the experimental observations, a given amount of lithium was allowed to deposit on the top of the porous Cu and a deposition gradient was applied along the depth from the top to the bottom. Then, the tortuosity factor was determined for each of the resulting structures using the software TauFactor,[33] which relies on Fickian diffusion. Because of the stochastic nature of the lithium deposition simulation, each simulation was repeated between 5 and 10 times to ensure reproducibility and sufficient statistics.

The tortuosity factor results are reported in Fig. 5a for each porous Cu samples as a function of the plated lithium capacity. It appears that the tortuosity factor follows a gentle exponential rise until a certain threshold where it sharply increases (see Supporting Information Fig. 5a and Table S3). The larger the pores, the higher the threshold capacity corresponding to the steep rise in the tortuosity factor. The 30Arc sample has the smallest pores, and as a result the tortuosity factor begins to increase rapidly at the lowest capacity for this sample (at around 7 mAh/cm$^2$). It is noteworthy that the 30Furnace sample shows a higher tortuosity factor than the 10Arc sample until a capacity of 7 mAh/cm$^2$; after that point, the two curves cross. From the 3D images on Fig. 5b, the large pores of the two structures do not look to be clogged with lithium at 4 mAh/cm$^2$. At low capacities, it thus seems that if the pores are large enough to accommodate the lithium deposition, the interconnectivity of the pores and the tortuosity of the empty pores will be the most important parameters. However, at 7 mAh/cm$^2$, the lithium deposits appear to be significantly denser. Indeed, beyond 7 mAh/cm$^2$ the 30Furnace structure displays a lower tortuosity factor which suggests that at higher capacities, the pore size becomes the driving parameter. These conclusions could assist in the optimization of the three-dimensional architecture of the porous copper anodes, depending on the application. For low capacities, one should put effort at tailoring the morphology of the pores, but for higher capacities, the limiting parameter appears to be the size of the pores so they are not blocked by lithium deposition.

Based on the observations above, the effect of the tortuosity factor can be summarized as following. The narrow and tortuous structures, such as the case in the 30Arc sample, would largely hinder the transport of lithium ions, leading to an inhomogeneous distribution of the deposited lithium. The inhomogeneous distribution would also waste the empty space and surface area provided by the 3D structure and possibly lead to the formation of lithium whiskers after the top surface is fully covered by the deposited lithium. Therefore, as mentioned previously, an increase in surface area does not always lead to a more uniform morphology for the deposited lithium. The increase of surface area and the decrease of pore size can have competing effects. From the data presented here, it appears that the sample with high enough surface area to provide a uniform morphology for lithium deposited at various plating rates and with pore that are large enough to facilitate lithium ion transport is the 20Arc material. The material seems to be near the "sweet spot" in this porous copper system.

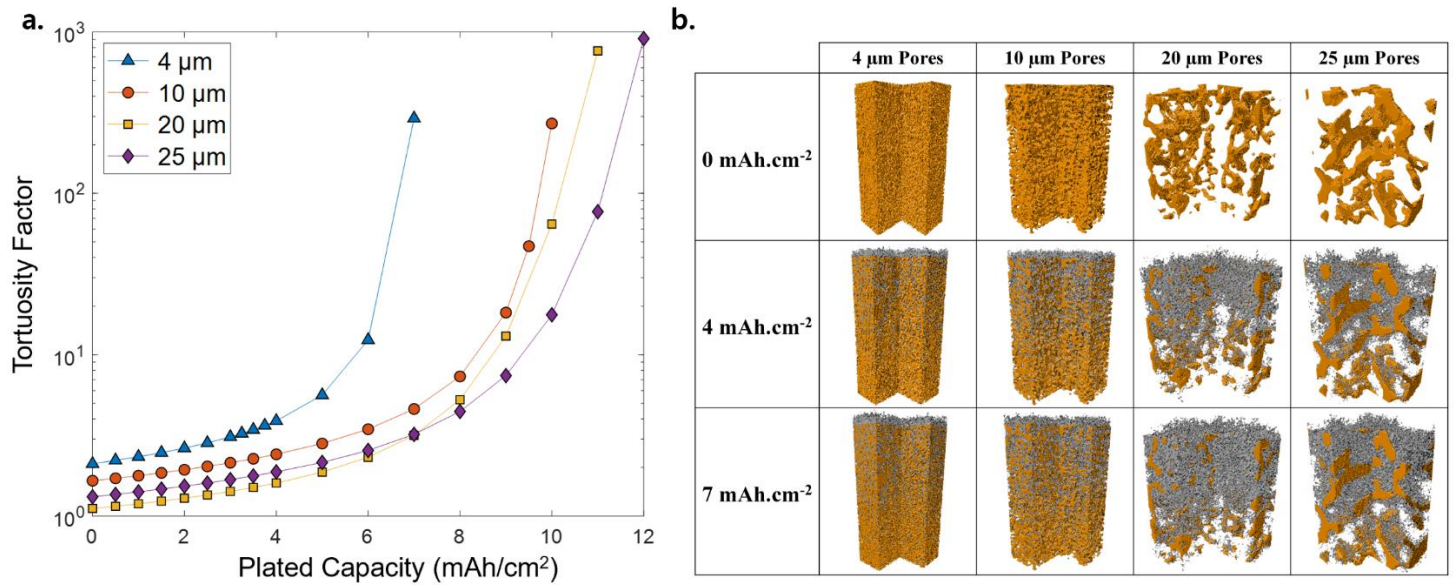

Figure 5. a) Tortuosity factor as a function of the capacity of Li plated into each porous copper structure; b) Porous coppers before (0 mAh/cm$^2$) and after (4 and 7 mAh/cm$^2$) Li deposition with the copper in gray and the Li in orange. The whole thickness of the current collectors is displayed with a square base of 100 µm of length. One quarter of each structure was removed for visualization-sake.

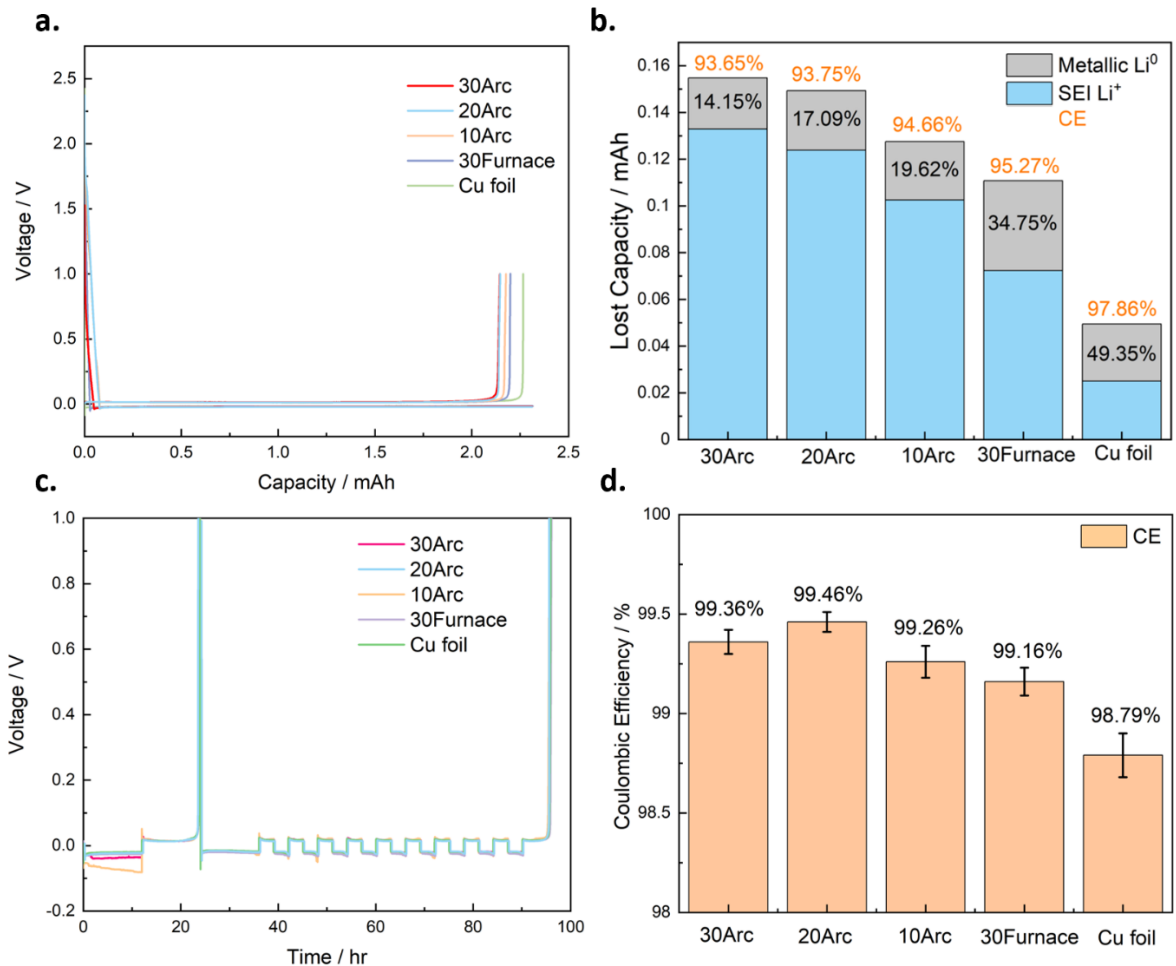

Figure 6. a) Electrochemical potential curve for lithium plating and stripping on each of the porous copper samples. b) TGC quantification of the SEI and inactive metallic lithium fractions for each sample. c) Electrochemical potential curve for CE testing for each porous copper sample. d) The calculated average CE of each porous copper sample from the half plating-stripping testing protocol.

To evaluate our prediction for the performance of the porous copper based on their physical parameters, Li||Cu cells were made to test the CE for lithium plating/stripping. The first cycle CE was calculated by plating 6mAh/cm$^2$ of lithium onto the porous coppers and stripping at 0.5mA/cm$^2$. After that, TGC was applied to quantify the amount of SEI Li$^+$ and inactive metallic Li$^0$ formed on the first cycle. In the TGC experiment, a fixed amount of deionized (DI) water is added to the cycled porous copper electrode, the inactive metallic lithium that was not removed during stripping will react with the DI water and release hydrogen gas. The amount the hydrogen gas quantified through gas chromatography can be directly correlated to the amount of the inactive metallic Li$^0$ left on the porous copper. It is known that CE values below 100% in each cycle of lithium plating/stripping comes from the combination of SEI Li$^+$ formation and formation of inactive metallic Li$^0$. By quantifying the fraction of inactive metallic Li$^0$, the SEI

Li$^+$ can then be calculated from the CE, and the combination of both values can help us to understand the failure model of each porous copper sample.

As shown in Fig 6b, the 30Arc samples showed the lowest CE on the first cycle, which was 93.65%. However, the amount of inactive metallic Li$^0$ formed was only 0.0219 mAh, which was only 14.15% of the total irreversible capacity, the lowest among the five types of coppers tested. The results show that although the high surface area of the 30Arc helped to improve the morphology of the deposited lithium, which is reflected in low percentage of inactive metallic Li$^0$, the high surface area also provide more contact surface between the electrolyte and lithium metal, which is thermodynamically unstable and leads to the formation of SEI under most conditions. In addition, during the first cycle due to the presence of CuO$_x$, Li will first react with this oxides layer and form SEI, which also contributes to the low CE. This is most prominently shown in the first cycle CE of the 30Arc sample, as it has the highest surface area. While the CuO$_x$ is an issue only in the first cycle, the SEI in the high surface area samples remains a problem throughout cycles. For the 20Arc samples, with the decrease in surface area, the amount of SEI also decreased while the formation of inactive metallic Li$^0$ remained low. Therefore, an increase in CE was observed. As the surface area further decreased, the formation SEI also decreased in the 10Arc and 30Furnace samples. However, due to the low surface area of the 30Furnace samples, some lithium whiskers formed in these samples (Fig. 3m), increasing the fraction of inactive metallic Li$^0$. The sample with the highest CE in the first cycle is the 2D copper foil. The high CE in the copper foil comes from two reasons: 1) The low surface area of the copper foil limits the contact area between lithium and electrolyte and the amount of CuO$_x$ that can form, thus leading to lower SEI formation; 2) The stack pressure on the copper foil from the coin cell also helped more lithium to be stripped back to the electrolyte, whereas the stacking pressure inside porous copper was nearly zero, so the lithium can more easily detach from the copper substrate during stripping.

Based on the analysis above, it can be concluded that although the high surface area can help lithium to grow in a better morphology, the extra surface area also causes the formation of extra SEI. These competing factors led us to use a more representative way to determine the CE of the lithium plating/stripping. As shown in Fig. 5c, a half plating and stripping method was utilized to characterize the CE of the lithium plating/stripping. In this method, first, a full lithium plating/stripping cycle was performed to condition the surface of the coppers. Then, 6 mAh/cm$^2$ of lithium reservoir (Q$_P$) was plated onto the coppers at 0.5 mA/cm$^2$, followed by 9 cycles of striping and plating of 3 mAh/cm$^2$ of the lithium (Q$_{half}$). At the last step, all the remaining lithium was stripped to the cutoff voltage of 1V (Q$_S$). The average CE is calculated by (9Q$_{half}$ + Q$_S$)/(9Q$_{half}$ + Q$_P$)×100%. While the conventional CE testing method can tell us the efficiency of each cycle, it will include the capacity lost due to the formation of fresh SEI on cycle. The half plating and stripping technique leaves a layer of lithium reservoir after the first cycle, which can lead the lithium to fill back into the pre-existing SEI layer to minimize its formation. Furthermore, the first formation cycle has also consumed most of the oxides layer on the porous copper, leaving behind a relatively clean surface for Li metal deposition. Using this method, the average CE of the 30Arc, 20Arc, 10Arc, 30Furnace and copper foil was calculated to be 99.36%, 99.46%, 99.26%, 99.16% and 98.79% as shown in Fig. 6d. The preconditioning and the half plating and stripping technique greatly minimized the formation of SEI in each cycle and the CE is directly reflecting effect of the morphology and the distribution of the deposited lithium. As predicted earlier, the 20Arc shows the highest CE because of its relatively high surface area and also large pore size.

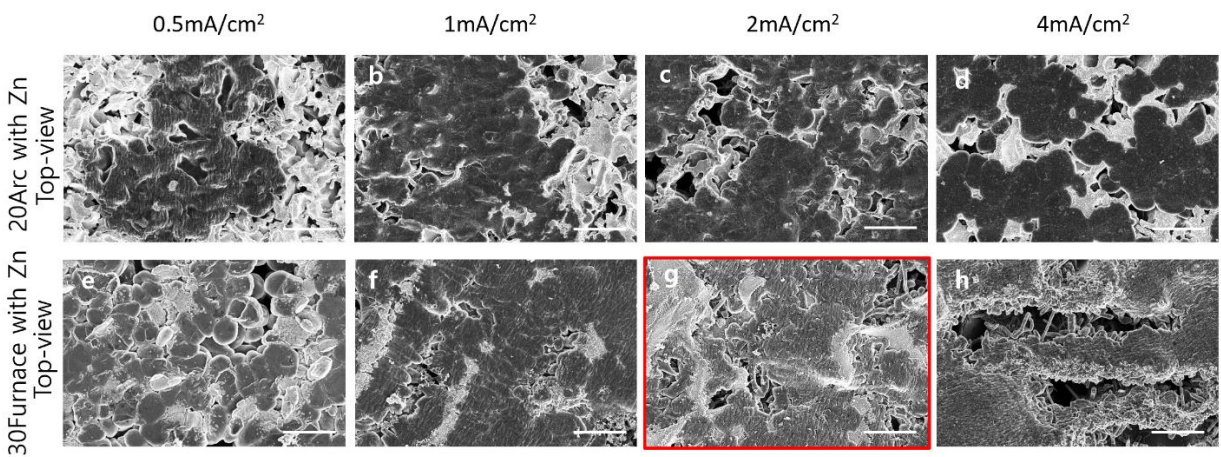

Figure 7. Top-view SEM images of the morphology of the lithium plated under different current density onto porous copper with a Zn coating. a-d) 20Arc sample coated with Zn; e-h) 30Furnace sample coated with Zn. The deposited Li is shown in darker contrast, while the brighter contrast shows the porous copper substrate. The critical current density at which lithium whiskers start to grow is marked with a red box. All samples were plated to 4mAh/cm$^2$.

The last parameter that can affect the performance of a 3D current collector is the surface chemistry. It is known that the electrochemical property of the substrate that Li is plated on can significantly influence the morphology of the Li.[18,19] To quantitatively explore how the improvement in surface chemistry can bring to the performance of the porous copper current collectors, metallic Zn, which has a lower nucleation barrier for Li than Cu,[19] was electrochemically coated onto 20Arc and 30Furnace samples. These two samples were chosen because they showed the best and the worst performance, respectively, based on the data presented above. The XRD in Figs. S10 and S11 show that upon Zn coating, the Cu and Zn form an alloy of $CuZn_5$ and $Cu_5Zn_8$. The SEM images of the as-made and post-coating samples (Figs. a and b in S12 and S14) show that the grains of the Cu-Zn alloy is in the sub-micron scale. Furthermore, the EDS in the coated samples (Fig. S13 and S15) shows coverage of zinc all over the sample with little oxygen.

To study the role of Zn coating, Li was electrochemically deposited onto the bare porous copper and Zn coated porous copper. The Li metal nucleation overpotential, which is defined as the voltage difference between the bottom of the voltage dip and the flat part of the voltage plateau, is a parameter that represents the energy barrier that Li needs to overcome to nucleate on the substrate.[19] As shown in Fig. S16, the nucleation overpotentials of 20Arc and 20Arc with Zn are found to be 46.1 mV and 32.5 mV respectively, meaning that the Zn coating successfully lower the nucleation barrier of Li depositing on the substrate, which should potentially lead to a more uniform and dense morphology.

The morphology of deposited Li on the Zn coated porous copper was then studied by SEM. As shown in Fig. 7a-d, the resulting Li morphology after plating in the 20Arc sample coated with Zn is dendrite free across the whole range of current density tested. This result stands in contrast to the uncoated 20Arc sample, where Li whiskers begin to appear at 4mA/cm$^2$. An even more dramatic improvement in Li morphology is observed in 30Furnace sample, however. As shown in Fig. 7e-h, the Li deposited Li on the 30Furnace sample coated with Zn shows a spherical morphology at 0.5 mA/cm$^2$. Even at this low current density, a whisker morphology was already seen in the uncoated samples. Li whiskers only begin to grow at a much higher current density of 2mA/cm$^2$.

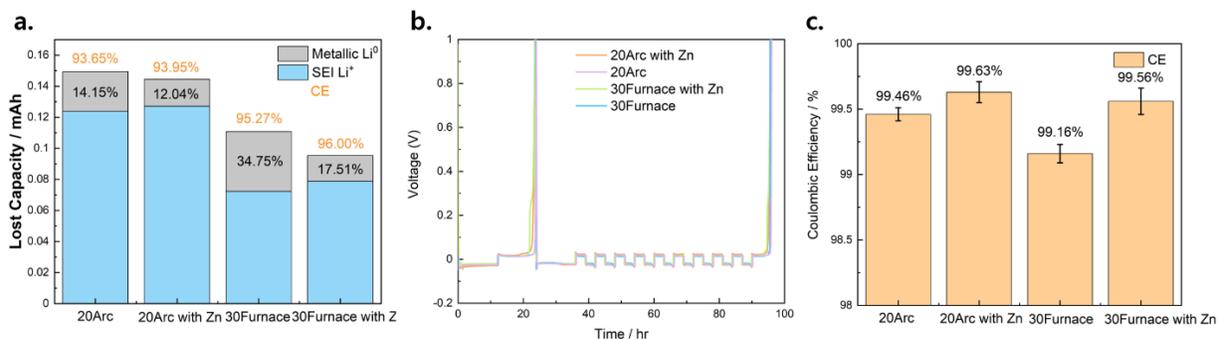

Figure 8. The electrochemical performance of Zn coated porous coppers: a) TGC analysis of the SEI and inactive metallic lithium; b) the electrochemical potential curve of CE testing of the porous copper samples; c) The calculated average CE of the Zn-coated porous copper sample from the half plating-stripping testing protocol.

The porous samples with Zn coatings were further tested for plating/stripping CE using the protocol mentioned previously. The 20Arc sample coated with Zn had an average CE of 99.63 %, which increased from 99.46 % of the uncoated samples, while the 30Furnace sample coated with Zn had a CE of 99.56 %, which increased from 99.16 % of the uncoated samples, as shown in Fig. 8c. The amount of SEI Li$^+$ and metallic Li$^0$ in the first cycled was again quantified using TGC (Fig. 8a). The first cycle CE of the 20Arc samples improved only a small amount after coating because the lithium morphology in the 20Arc was already dense without the coating, and the surface area was not significantly changed by the coating so that the SEI formed from the reaction between Li and oxides layer did not change much. The absolute amount of inactive metallic Li$^0$ did markedly decrease, however, presumable because of more effective Li nucleation on the improved surface. This was partly offset by a slight increase in the SEI Li$^+$, likely because the Cu$_4$Zn is more reactive toward oxygen or water than pure Cu. A more significant improvement was observed in the 30Furnace after coating. The first cycle CE improved from 95.27 % to 96.00 %, while the amount of inactive metallic Li$^0$ also markedly decreased because of the improved morphology. Again a slight increase in the SEI Li$^+$ was observed, but change was not large enough to offset the large gains from the reduction in inactive metallic Li$^0$ due to the improved plating onto the lithophilic surface. The decrease in the inactive metallic Li$^0$ and improved cycling performance was more obvious in the 30Furnace sample upon coating with Zn than in the 20Arc sample, which proved that tuning the surface chemistry is an effective way to enable a 3D current with low surface area to achieve uniform Li plating.

**Conclusions:** In this work, porous copper scaffolds with interconnected micro-channels were synthesized using wet chemical etching and used as 3D current collectors for lithium metal anode. Laboratory X-ray tomography was used to quantify the physical properties of the porous copper materials. The effects of the three key parameters – surface area, tortuosity, and surface chemistry – were carefully studied. Contradicting the conventional believe, we found that the highest surface area was harmful to the performance of the 3D current collectors, as Li metal was not able to plate deep into the pores due to diffusion limitations and more SEI was formed on the first cycle. A moderate surface area is desirable for a 3D current collectors where the local current density will be low enough to suppress the growth of lithium whiskers while keeping the formation of SEI to minimum. The tortuosity of the porous copper mainly influences the diffusion of lithium ions. To facilitate a uniform distribution of deposited lithium across the whole 3D current collector, the pore size should be kept as large as possible while the tortuosity should be minimized. In considering all of these key physical parameters, the 20Arc samples, with a balance of modest surface area and reasonable pore size, was shown to have the best performance among the copper samples tested, while the 30Furnace sample, which has very large pores and low surface area, had the worst.

    These two samples were further coated with metallic Zn to evaluate how surface chemistry can influence the performance of 3D current collectors. Upon electrochemical testing, an average lithium plating/stripping CE of 99.56% was achieved in 30Furnace with Zn coated, which improved from 99.16% without coating. However, the CE for the 20Arc sample did not show as much improvement with the Zn coating because the morphology of the Li in 20Arc was already exceptionally good without coating. The results in this work led to several key points: 1) Although the high surface area can lower the local current density and lead to a uniform Li morphology, it will also introduce large amount of SEI formation because of the increase surface area. 2) The tortuosity of the 3D current collector should be kept at minimum to induce the uniform distribution of Li in the structure. 3) The Zn coating can effectively improve the morphology of the Li, allowing even 3D current collectors with low surface area to achieve uniform Li plating.

Citation

**Acknowledgments**:

**Funding**: This work was supported by the Office of Vehicle Technologies of the U.S. Department of Energy through the Advanced Battery Materials Research (BMR) Program (Battery500 Consortium) under Contract DE-EE0007764. SEM was performed at the San Diego Nanotechnology Infrastructure (SDNI), a member of the National Nanotechnology Coordinated Infrastructure, which is supported by the National Science Foundation (grant ECCS-1542148). AAF and MC acknowledge the European Union's

Horizon 2020 research and innovation programme for the funding support through the European Research Council (grant agreement 772873, "ARTISTIC" project). A.A.F. acknowledges Institut Universitaire de France for the support. This work was performed in part at the San Diego Nanotechnology Infrastructure (SDNI) of UCSD, a member of the National Nanotechnology Coordinated Infrastructure, which is supported by the National Science Foundation (Grant ECCS1542148).

**Supporting Information**

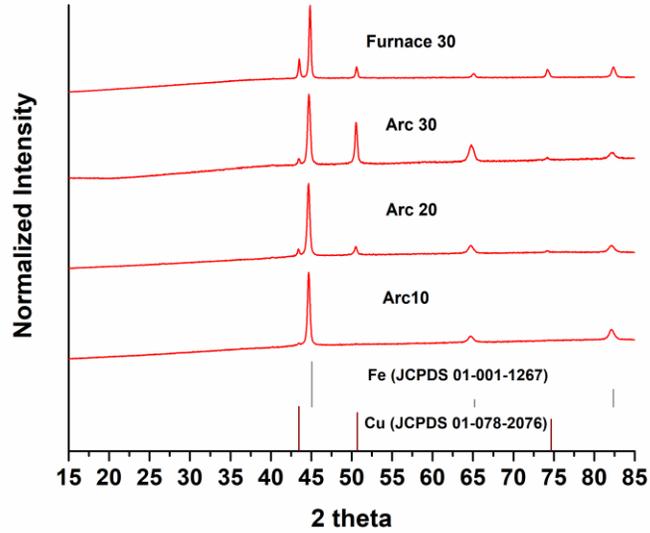

**Figure S1.** X-ray diffraction of the four Cu-Fe composites used in this work: 10Arc, 20Arc, 30Arc and 30Furnace. Also included are reference patters for Cu and Fe metal. All samples are composed of a combination of pure Fe and pure Cu.

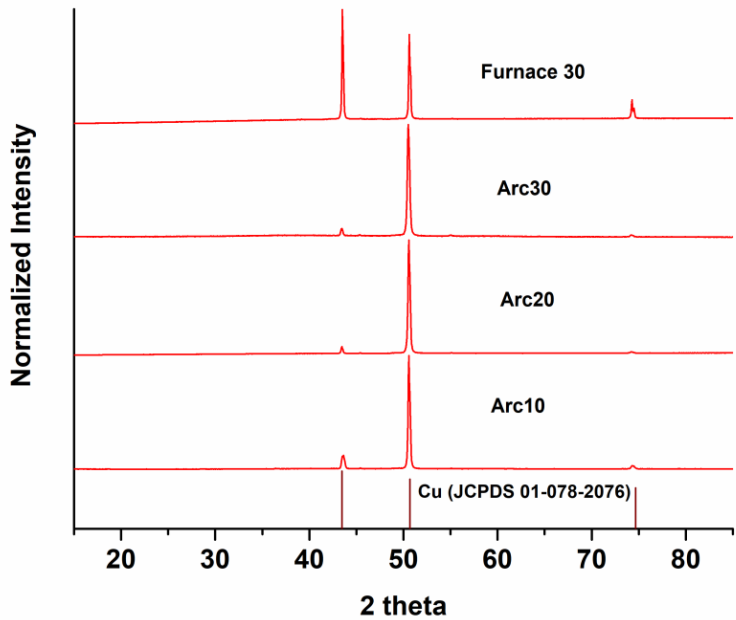

**Figure S2.** X-ray diffraction patterns of the four porous Cu samples used in this work after acid etching and cleaning. A Cu reference pattern is also included. The sample prepared using a furnace derived Cu-Fe composites showed minimal preferential grain orientation, while the arc melted samples showed significant preferred orientation.

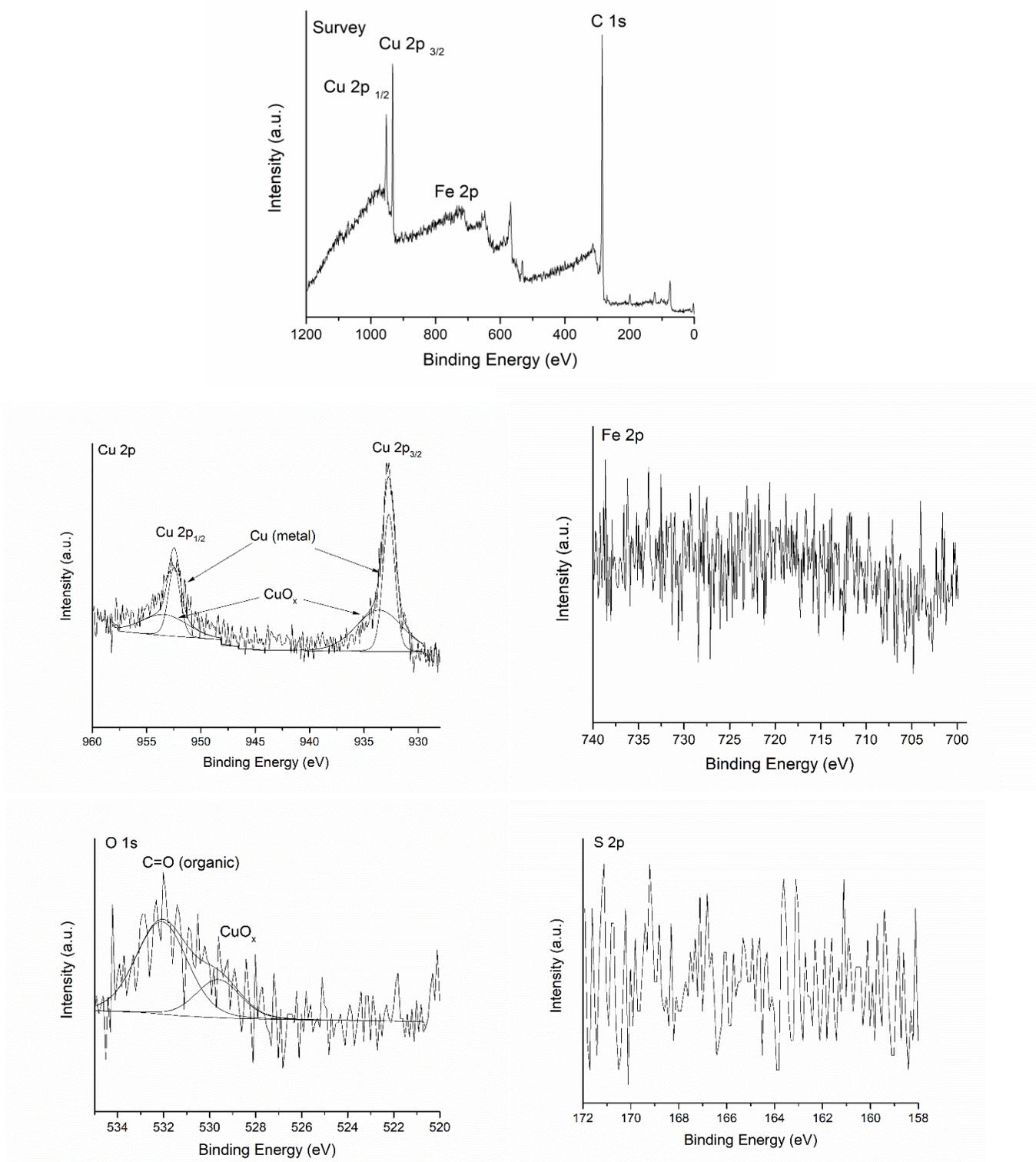

**Figure S3.** XPS of the 30Furnace sample after acid etching and cleaning. The sample was etched using argon ion sputtering at 4kV for 1min. The sample is composed of Cu and some surface $CuO_x$, with no Fe or S detected.

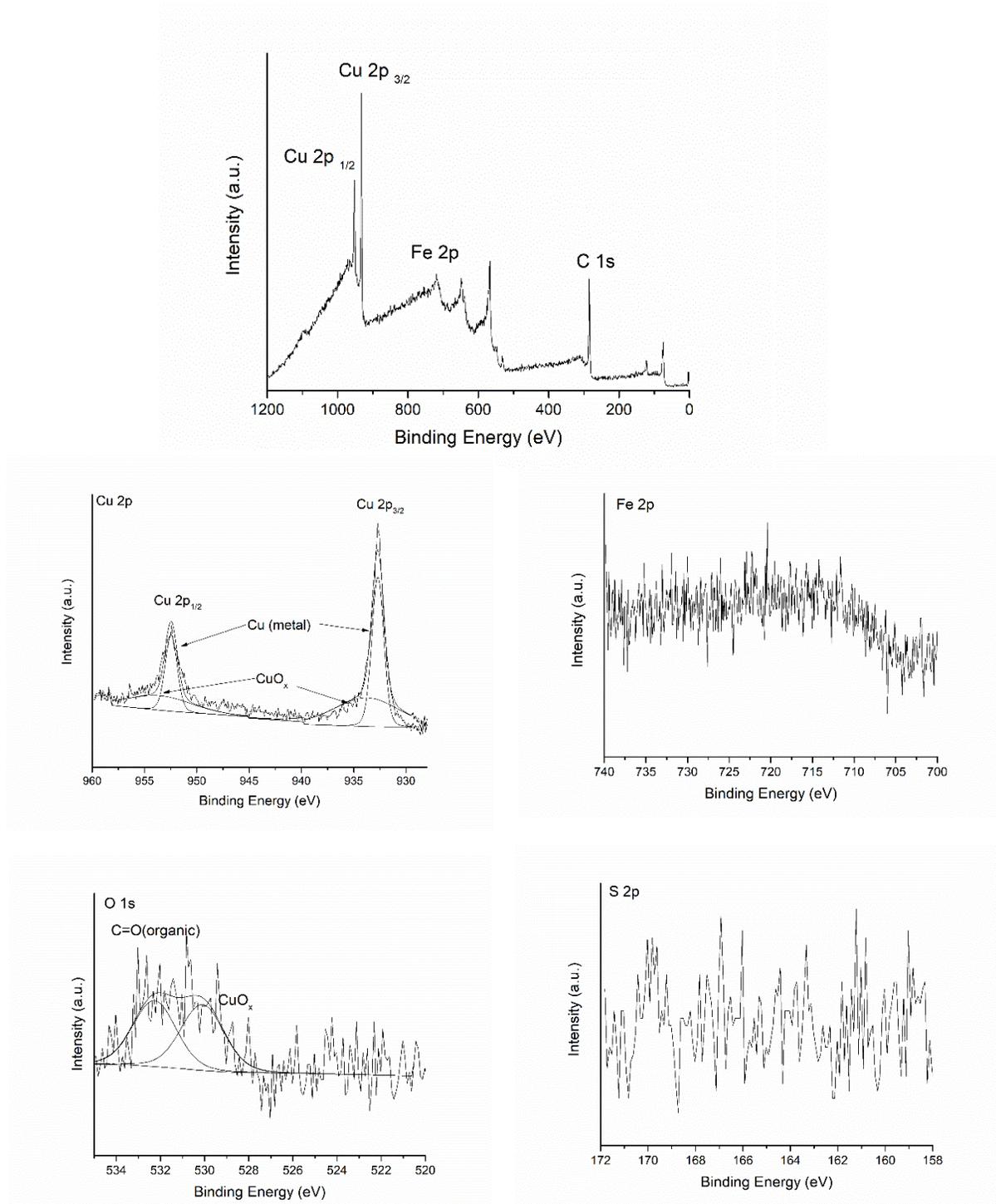

**Figure S4.** XPS of the 30Arc sample after acid etching and cleaning. The sample was etched using argon ion sputtering at 4kV for 1min. The sample is composed of Cu and some surface $CuO_x$, with no Fe or S detected.

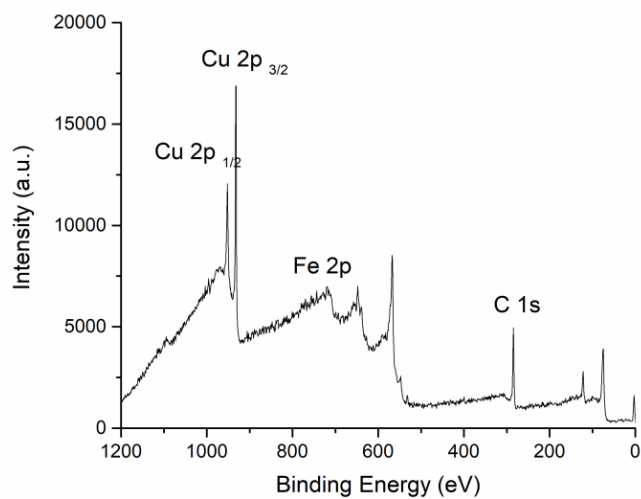

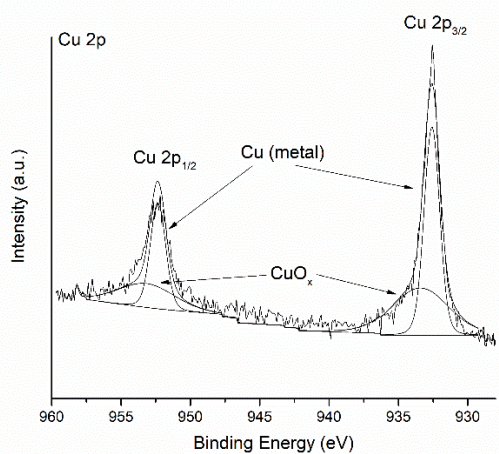
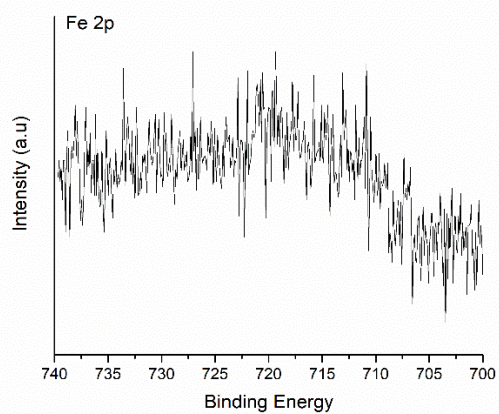

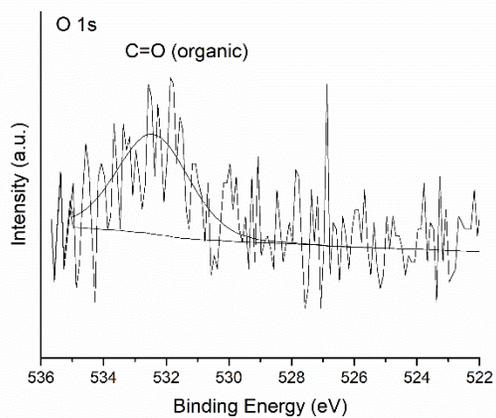
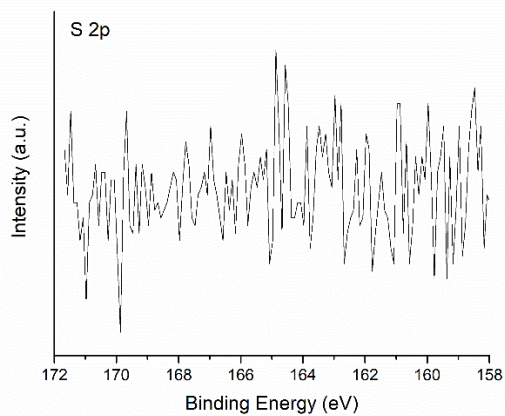

**Figure S5.** XPS of the 20Arc sample after acid etching and cleaning. The sample was etched using argon ion sputtering at 4kV for 1min. The sample is composed of Cu and some surface $CuO_x$, with no Fe or S detected.

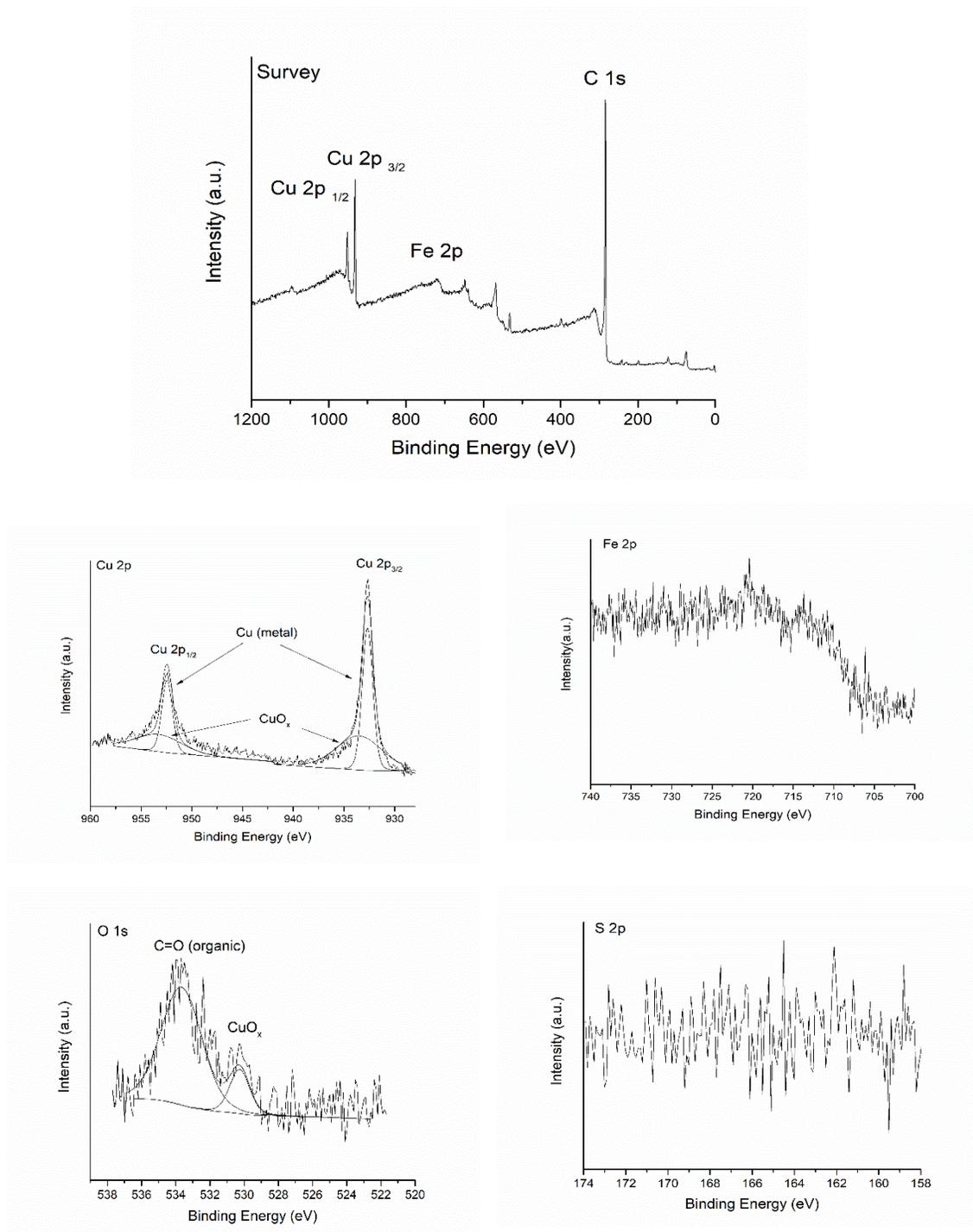

**Figure S6.** XPS of the 10Arc sample after acid etching and cleaning. The sample was etched using argon ion sputtering at 4kV for 1min. The sample is composed of Cu and some surface $CuO_x$, with no Fe or S detected.

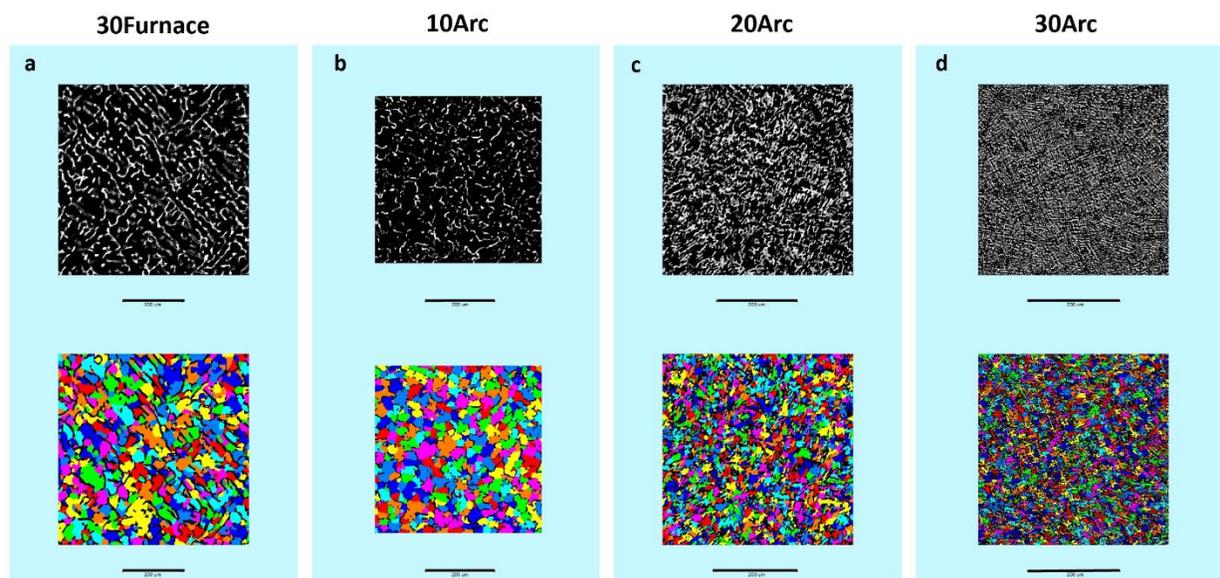

**Figure S7.** Examples of 2D slices from the X-ray tomography study and the separation of the image into pores. Samples shown are a) 30Furnace, b) 10Arc, c) 20Arc, and d) 30Arc.

| Sample | Li deposition rate on top |
|---|---|
| 30Arc | 15 % |
| 20Arc | 10 % |
| 10Arc | 5 % |
| 30Furnace | 5 % |

**Table S1.** Percentage of Li deposited on top of the copper current collector for each porous Cu sample.

| Copper Height (%) | Li deposition rate in the foam |
|---|---|
| 0 % → 16.6 % | 33 % |
| 16.6 % → 33.3 % | 27 % |
| 33.3 % → 50 % | 18.75 % |
| 50 % → 66.6 % | 10.5 % |
| 66.6 % → 83.3 % | 6.25 % |
| 83.3 % → 100 % | 4.5 % |

**Table S2.** Percentage of Li deposited in each sub-volume for the porous copper sample. Here, 0 % in thickness corresponds to the porous Cu surface closest to the separator.

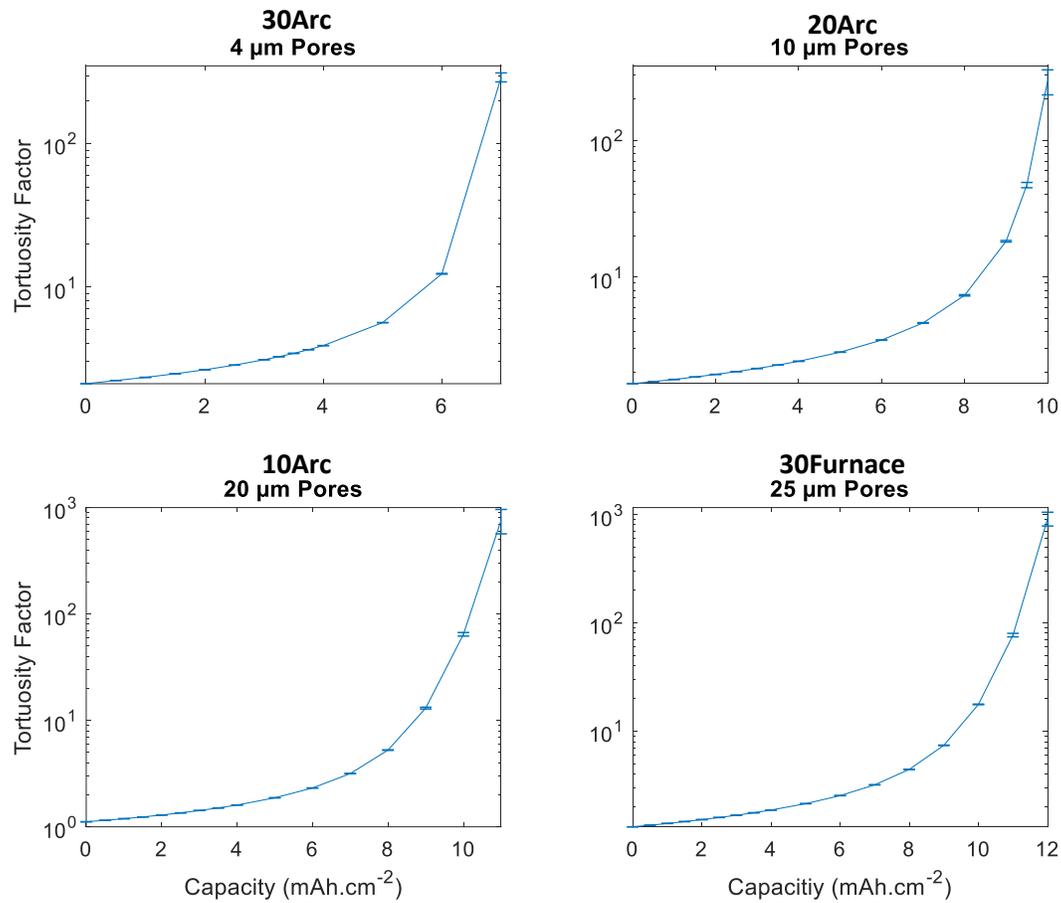

**Figure S8.** Tortuosity factor as a function of capacity with the associated error for each current collector. Samples names are listed on the figure.

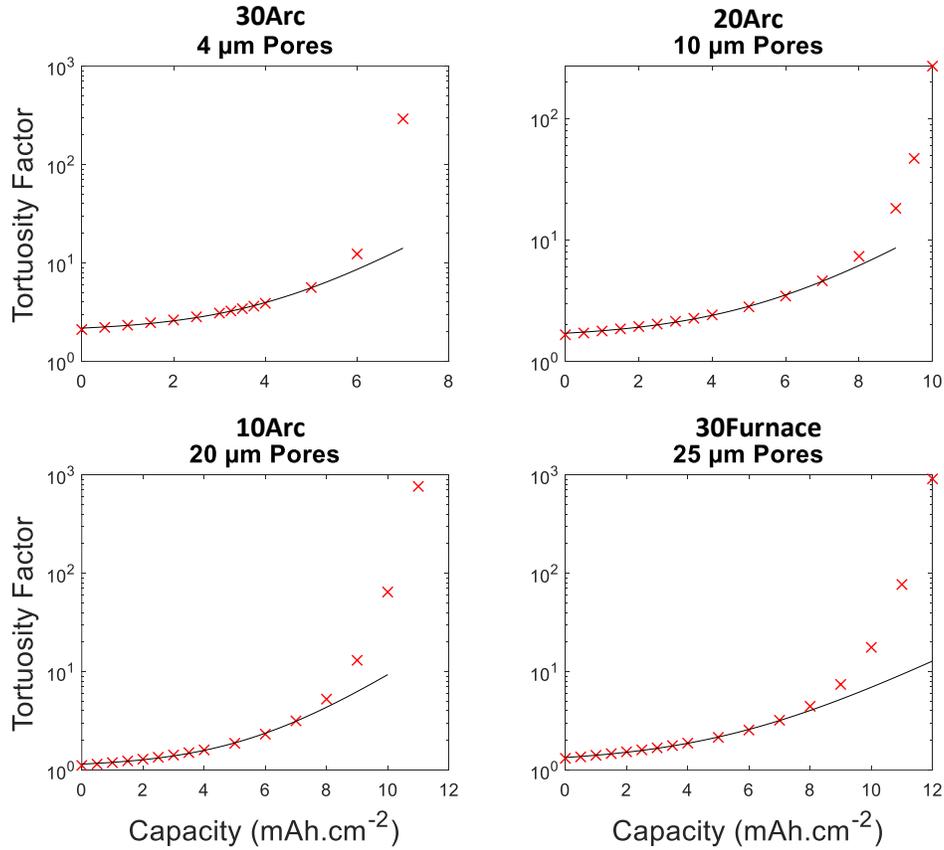

**Figure S9.** The evolution of the tortuosity factor as a function of the capacity was fit using Eq. S1.

$$T(c) = \alpha_0 + \alpha_1 e^{\alpha_2 c} \qquad (Eq.S1)$$

Only the region before the sharp rise in tortuosity, *i.e.* 5 mA·cm$^{-2}$ for the 4 µm pores and 7 mA·cm$^{-2}$ for the rest, was used for the fits; the fit parameters can be found in Table S3. The second region could not be fitted using the same function. In the figure, the tortuosity factor from the simulations (same data as figure S8) is shown in red and the associated exponential fits to the data before the steep increase in tortuosity are shown with a solid black line. Samples names are listed on the figure.

|  | $\alpha_0$ | $\alpha_1$ | $\alpha_2$ | $R^2$ |
|---|---|---|---|---|
| Arc30 | 2.0086 | 0.1702 | 0.6096 | 0.9980 |
| Arc20 | 1.5513 | 0.1555 | 0.4239 | 0.9981 |
| Arc10 | 1.0696 | 0.0816 | 0.4619 | 0.9979 |
| Furnace30 | 1.1700 | 0.1730 | 0.3509 | 0.9988 |

**Table S3.** Fit parameters from the function given in Eq.S1 and the associated coefficients of determination $R^2$.

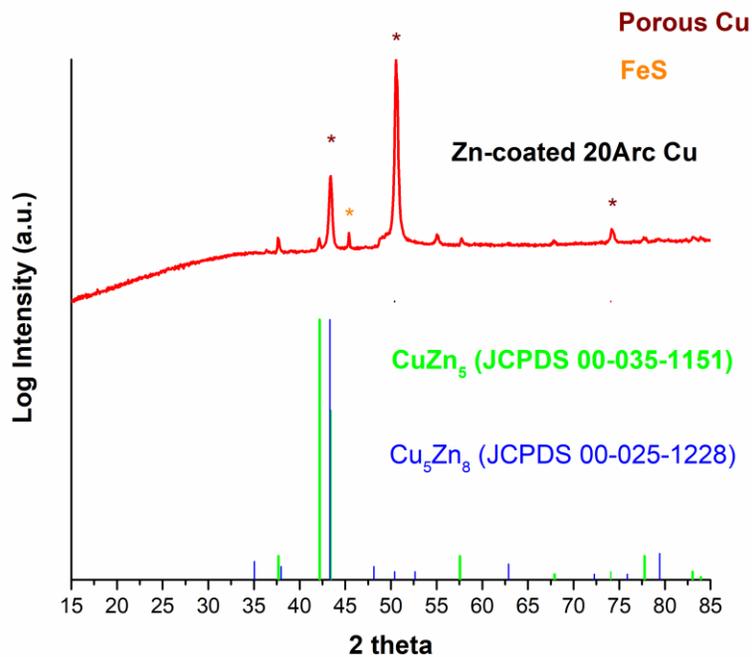

**Figure S10.** X-ray diffraction pattern of the Zn-coated 20Arc Cu sample after dealloying, coating, and cleaning. Diffraction peaks belonging to Cu metal for FeS are indicated with *.

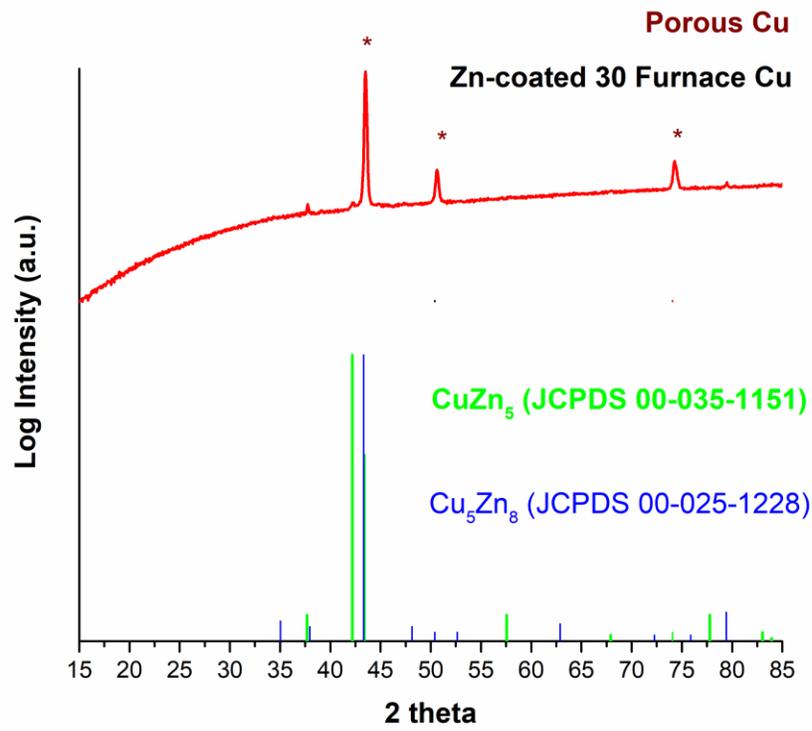

**Figure S11.** X-ray diffraction pattern of the Zn-coated 30Furnace Cu sample after dealloying, coating, and cleaning. Peaks corresponding to pure Cu are marked with *.

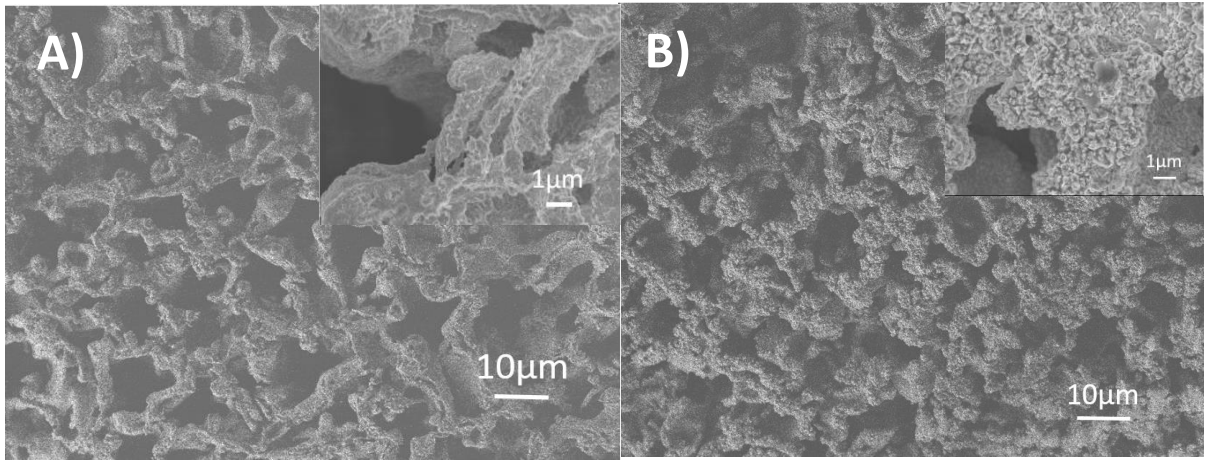

**Figure S12.** Top-view SEM images of the A) as made-20Arc Cu and B) Zn-coated 20Arc Cu samples. The images look very similar indicating the formation of a conformal coating.

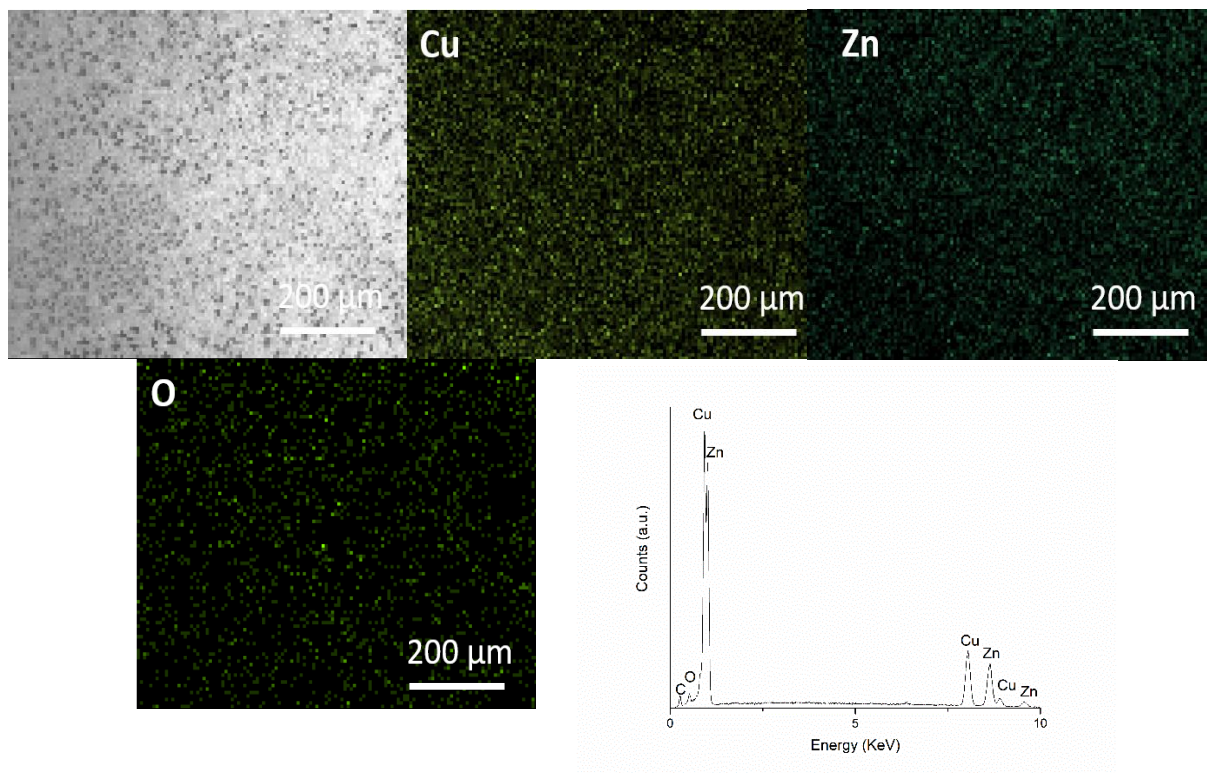

**Figure S13.** EDS mapping for the 20Arc Cu sample after Zn deposition showing the maps of Cu, Zn, O and the overall EDS energy spectrum. Uniform distributions of Cu and Zn indication a homogeneous coating.

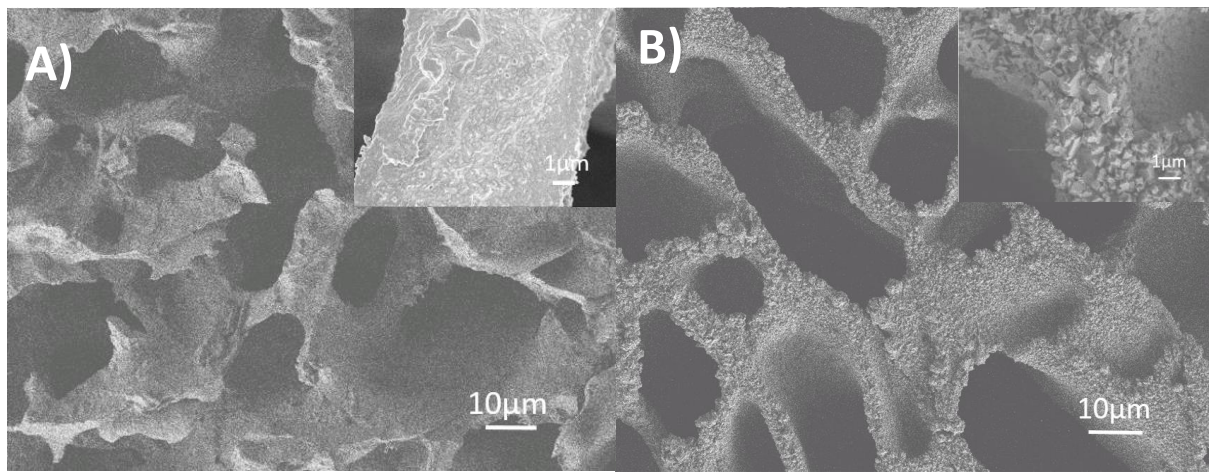

**Figure S14.** Top-view SEM images of the A) as made-30Furnace Cu and B) Zn-coated 30Furnace Cu samples. The images look very similar indicating the formation of a conformal coating.

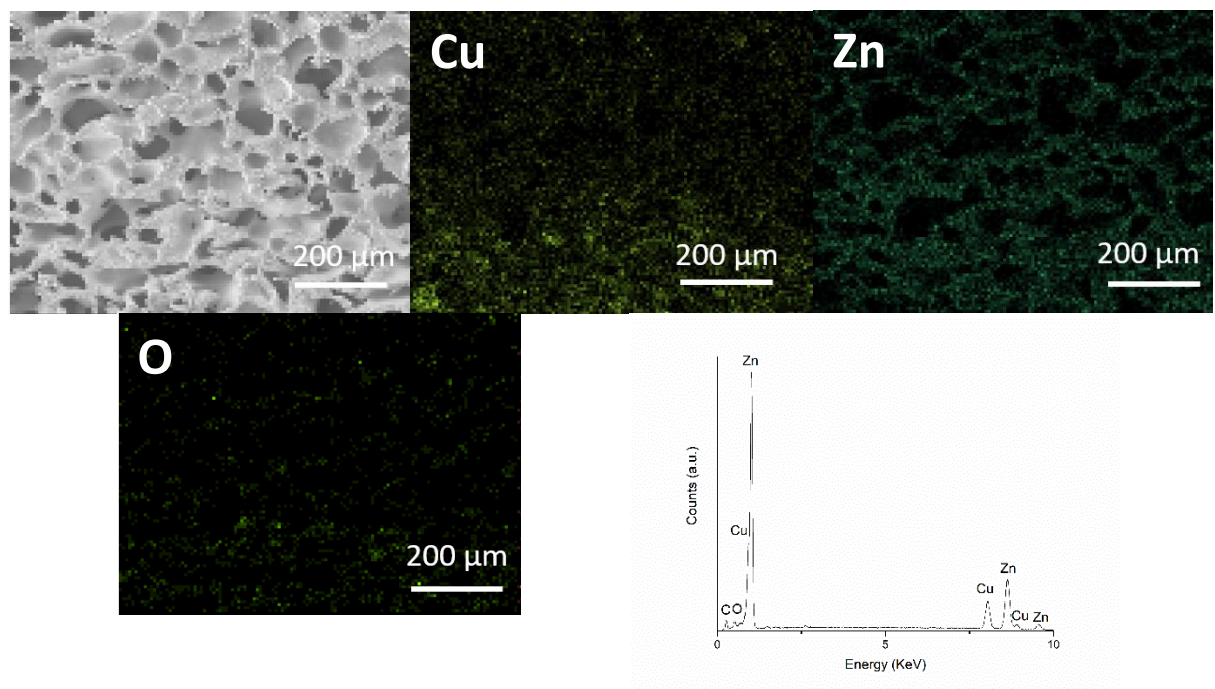

**Figure S15.** EDS mapping for the 30Furnace Cu sample after Zn deposition showing the maps of Cu, Zn, O and the overall EDS energy spectrum. Uniform distributions of Cu and Zn indication a homogeneous coating.

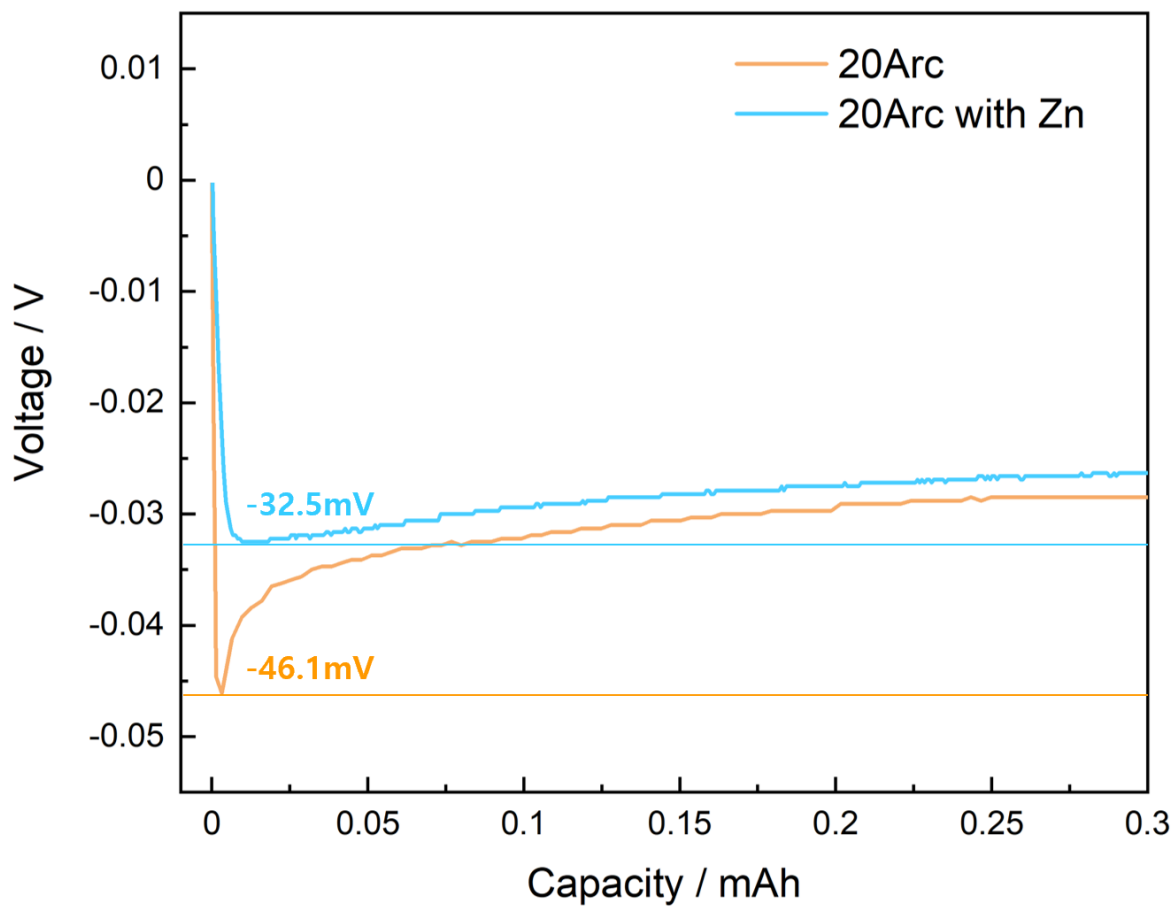

**Figure S16.** The voltage profile of Li deposition on the 20Arc and the Zn-coated 20Arc sample. The nucleation overpotential is labelled with a horizontal line for each sample. Zn coating results in a significant decrease in nucleation overpotential.